\documentclass[preprint,showpacs,preprintnumbers,amsmath,amssymb,nofootinbib,showkeys]{revtex4}

\usepackage{bm}
\usepackage{subeqnarray}
\usepackage{mathrsfs}
\usepackage{dsfont}

\begin{document}


\title{Generalization of Gibbs Entropy and Thermodynamic Relation}

\author{Jun Chul Park}
\email{parkjunchul@hotmail.com}
\affiliation{A 201, 362-5 Mokcheon, Cheonan, SEOUL 330-844 Republic of KOREA}

%
%


\begin{abstract}
In this paper, we extend Gibbs's approach for quasi-equilibrium
thermodynamic processes, and  show that, in general non-equilibrium
thermodynamic processes, the microscopic expression of entropy is
given as
\begin{equation}
S(t)=-\int dx  \rho (x,t) \ln \int d x' \rho (x',t)
\phi_{\mathit{\Delta} t} (x, x',t), \nonumber
\end{equation}
where $\rho (x,t)$ is the ensemble distribution in phase space and
$\phi_{\mathit{\Delta} t} (x,x',t)$ is the probability density to
obtain that, in macroscopic observation, the system with initial
value $x'$ in phase space at time $t$ is found at state $x$ after
time elapse $\frac{\mathit{\Delta} t}{2}$, and $\mathit{\Delta} t$
is the maximum value of the time interval for which any macroscopic
thermodynamic variables increase linearly.  Also, we analyze the
formal structure of thermodynamic relation in non-equilibrium
thermodynamic processes.
\end{abstract}

\pacs{05.20.-y, 05.20.Gg}
\keywords{Non-equilibrium entropy; Thermodynamic relation}
\maketitle

Since Gibbs found the microscopic expression of entropy in the
course of analyzing the quasi-equilibrium thermodynamic process
\cite{Gibbs,Einstein},  there has been much effort to formulate the
generalized theory for entropy to be applicable in non-equilibrium
thermodynamic processes.   Although recent progress related to the
fluctuation theorem gives some information necessary to obtain such
generalization \cite{Evans}, it seems still
far from obtaining an adequate foundation for the general theory of
entropy, in that we have not yet gained a clear connection between
entropy and microdynamics in non-equilibrium processes.   The purpose
of this paper is to develop a consistent theory to connect entropy
and the classical dynamics in general non-equilibrium processes.
In particular, it will be shown that the developed theory  overcomes
the conventional problem as the Gibbs entropy cannot discriminate
the irreversible variation of a closed Hamiltonian dynamical system
as it is an invariant for such a system (by Liouville's theorem),
which is important in constructing a systematic theory of arbitrary
non-equilibrium thermodynamic processes in future.    In this work, we
obtain the microscopic expression of entropy in general
non-equilibrium thermodynamic processes, and prove the entropy theorem
for an arbitrary closed (ergodic and chaotic) Hamiltonian dynamical
system. Also, we will obtain the extended thermodynamic relation for
general thermodynamic processes and derive the fluctuation theorem.

\section{Complex dynamical system interacting with external environment}
Let us consider the Hamiltonian system
\begin{equation}
\mathscr{H}= H(q,p) +V_{\text{int}}(q,\pi), \label{Hamiltonian}
\end{equation}
where the Hamiltonian system $H$ defined by the coordinates $q
\equiv (q_1, q_2, \dots, q_N)$ and momenta $p \equiv (p_1, p_2,
\dots, p_N)$ interacts with its environment system, which is
described by the coordinates $\pi \equiv (\pi_1,
\pi_2,\dots,\pi_{\nu})$, by the interaction potential
$V_{\text{int}}(q,\pi)$. If there is macroscopic slow motion of
$\pi$ denoted by the time series $a(t) \equiv (a_1(t),
a_2(t),\dots,a_{\nu}(t))$, we define new variables as $\pi_i'=\pi_i -
a_i(t)$. In macroscopic point of view, the interaction by $a_i(t)$
is still interpreted as a mechanical interaction; on the contrary,
the interaction by $\pi_i'$ is the quantity interpreted as thermal
interaction.  Instead of using the variables $\pi_i'$, we replace
the dynamical dependence of $\pi_i'$ by the dependence of time $t$,
treating $\pi_i'$ as a fast fluctuating time series, and the Hamiltonian
is written as
\begin{equation}
\mathscr{H}= H(x) +V_{\text{int}}(q,a,t),
\end{equation}
where $x \equiv (x_1, x_2,\dots,x_{2N}) \equiv (q_1, \dots, q_N,
p_1, \dots, p_N) \equiv (q,p)$.  Then, the probability density
function (PDF) $\rho(x,t)$ describing the ensemble distribution in
the phase space $\mathit{\Gamma}$ defined by $x$ satisfies the
Liouville equation:
\begin{equation}
\frac{\partial \rho (x,t)}{\partial t}= -i L(x,a,t) \rho (x,t),
\label{Liouville's equation}
\end{equation}
where $i L(x,a,t) \equiv \sum_{k} \frac{\partial
\mathscr{H}}{\partial p_k}  \frac{\partial}{\partial q_k}- \sum_{k}
\frac{\partial \mathscr{H}}{\partial q_k} \frac{\partial}{\partial
p_k}$, and, by the normalization condition for $\rho (x,t)$,
\begin{equation}
\int  \rho (x,t) dx = 1, \label{norm}
\end{equation}
where $dx \equiv dx_1  dx_2\cdots dx_{2N}$.

Let the characteristic time interval for our macroscopic measurement
system be $\mathit{\Delta} t$ in the sense that the error $\xi_0$ of
macroscopic time-measurement is given within the error range
\begin{equation}\label{time error range}
-\frac{\mathit{\Delta} t}{2} <\xi_0 < \frac{\mathit{\Delta} t}{2}
\end{equation}
approximately. In such a case, \emph{the experimentally discriminable
minimum of time elapse from a given instant is
$\frac{\mathit{\Delta} t}{2}$ in statistical sense}. Then, the time
series $a_i(t)$ can be interpreted as a time-averaged behavior of
$\pi_i(t)$ during the error range $\mathit{\Delta} t$ such as
$\frac{1}{\mathit{\Delta} t} \int_{t}^{t+\mathit{\Delta} t}\pi_i(t')
dt'$, and we reasonably request that, if there is increment
$\mathit{\Delta} a \equiv (\mathit{\Delta} a_1, \mathit{\Delta} a_2,
\dots, \mathit{\Delta} a_{\nu})$ in $a$ during $\mathit{\Delta} t$,
$\mathit{\Delta} a_i$ is a linear function of $\mathit{\Delta} t$:
\begin{equation}\label{condition for a0}
\mathit{\Delta} a_i \propto \mathit{\Delta} t.
\end{equation}
Generally, since arbitrary macroscopic slow variation in system
$\mathscr{H}$ can be interpreted as such a time-averaged behavior of
some complex dynamical variation during $\mathit{\Delta} t$,
\emph{$\mathit{\Delta} t$ is the maximum time interval during which
we can approximate that any macroscopic thermodynamic variable varies
linearly for time}.

On the other hand, based on the dynamic complexity of system
$\mathscr{H}$, for the experimentally discriminable minimum time
elapse $\frac{\mathit{\Delta} t}{2}$, we have the following
presupposition: since we assume that, including the case where the
system is completely isolated from the external variation as $\pi_i
= \mspace{-1mu}\text{const}$, there are always some fast chaotic
and ergodic motions such that their trajectory varies non-linearly
during $\frac{\mathit{\Delta} t}{2}$, we have the condition
\begin{equation}
\hat{T}  e^{-i \int_{t}^{t + \frac{\mathit{\Delta} t}{2}}L(x,\tau)
d\tau}  \not \approx 1 -i L(x,a,t) \frac{\mathit{\Delta} t}{2},
\label{fastmotion condition}
\end{equation}
where $L(x, \tau) \equiv L(x,a(\tau),\tau)$ and $\hat{T}$ denotes
the time-ordering operator which interchanges the operators to
follow in chronological order from right to left; if the system is
completely isolated, it is written as $e^{-i L(x,a)
\frac{\mathit{\Delta} t}{2}} \not \approx 1 -i L(x,a)
\frac{\mathit{\Delta} t}{2}$.

Similarly, if the characteristic interval of $x$ for the given
macroscopic measurement system is $\mathit{\Delta} x\equiv
(\mathit{\Delta} x_1, \mathit{\Delta} x_2, \dots, \mathit{\Delta}
x_{2N})$ in the sense that the error $\xi \equiv (\xi_1, \xi_2,
\dots,\xi_{2N})$ of the macroscopic measurement for $x \equiv (x_1,
x_2,\dots,x_{2N})$ satisfies the error range
\begin{equation}\label{coordinates error range}
-\frac{\mathit{\Delta} x_i}{2} <\xi_i < \frac{\mathit{\Delta}
x_i}{2}
\end{equation}
approximately, we assume that, including the case of the completely isolated system,
there are always some complex (chaotic and ergodic) motions
macroscopically undescribable as a smooth path in the sense that, by these complex motions,  the velocity
field in $\mathit{\Gamma}$ or the Liouville operator $L(x,a,t)$ cannot be approximated
linearly for the increment $\frac{\mathit{\Delta} x_i}{2} $ in $x$; therefore we have
\begin{equation}
L( x_i + {\textstyle \frac{\mathit{\Delta} x_i}{2}}, a, t) \not
\approx L (x,a,t)+ \frac{\partial L(x,a,t)}{\partial x_i}
\frac{\mathit{\Delta} x_i}{2}. \label{fastmotion condition2}
\end{equation}
%
%

In the following, we will trace the rational consequence of the
presupposition (\ref{time error range}),  (\ref{condition for a0}),
(\ref{fastmotion condition}), (\ref{coordinates error range}), and
(\ref{fastmotion condition2}).

\section{Macroscopic behavior of ensemble distribution}
For the macroscopic system characterizing such values
$\mathit{\Delta} t$ and $\mathit{\Delta} x$, the experimentally
observed quantity for $\rho (x,t)$ is given by $\tilde{\rho} (x,t)$
as
\begin{equation}
\tilde{\rho} (x,t) = \frac{1}{\mathit{\Delta} t  \mathit{\Delta}
\mathit{\Gamma}} \int_{t}^{t+\mathit{\Delta} t}\int_{x}^{x+ \mathit{\Delta}
x}dt' dx'  \rho (x',t'), \label{definition of rho bar}
\end{equation}
where $\mathit{\Delta} \mathit{\Gamma}\equiv \mathit{\Delta} x_1
 \mathit{\Delta} x_2 \cdots \mathit{\Delta}
x_{2N}$ is the volume of
$\mathit{\Delta} x$; or more precisely
\begin{equation}
\tilde{\rho} (x,t) = \int_{- \infty}^{\infty} \int_{-
\infty}^{\infty} d \xi_0  d \xi  f(\xi, \xi_0)  \rho (x + \xi, t+
\xi_0) , \label{definition2 of rho bar}
\end{equation}
where $f(\xi, \xi_0)$ is the PDF to obtain the errors $\xi$ and
$\xi_0$ in the macroscopic measurement. The slowly varying property of
$\tilde{\rho} (x,t)$, which is implicitly understood from the
definition (\ref{definition of rho bar}) and (\ref{definition2 of rho bar})
based on the complexity of the chaotic and ergodic dynamics generating
the results (\ref{fastmotion condition}) and (\ref{fastmotion condition2}),
can be expressed as
\begin{subequations}\label{condition1 for rho bar}
\begin{align}
\tilde{\rho} (x,t + \mathit{\Delta} t) &\approx \tilde{\rho} (x,t)+
\frac{\partial \tilde{\rho} (x,t)} {\partial t} \mathit{\Delta} t,
\label{condition-time for rho bar}\\
\tilde{\rho} ( x_i + \mathit{\Delta} x_i, t)&\approx  \tilde{\rho}
(x,t) + \frac{\partial \tilde{\rho} (x,t)}{\partial x_i}
\mathit{\Delta} x_i
\end{align}
\end{subequations}
[in other words, for the chaotic and ergodic trajectories satisfying
(\ref{fastmotion condition}) and (\ref{fastmotion condition2}),
it will be assumed that the related complex dynamics is given so as
to make the averaged macroscopic quantity $\tilde{\rho} (x,t)$
acquire the property (\ref{condition1 for rho bar})].
Meanwhile, by the normalization condition for $\tilde{\rho} (x,t)$,
\begin{equation}
\int  \tilde{\rho} (x,t) dx = 1. \label{norm for rho bar}
\end{equation}

The slowly varying property (\ref{condition1 for rho bar}) of
$\tilde{\rho} (x,t)$ means that, for a
given thermodynamic process, $\tilde{\rho} (x,t)$ is expressible by
a smaller set of independent physical observables $\epsilon \equiv
(\epsilon_1,\epsilon_2,\dots,\epsilon_{\mathscr{N}})$ than $x$,
where $\epsilon_i$ is defined by $x$ and $a$, i.e.,
$\epsilon=\epsilon(x,a)$; for example, the energy $H(x) +
V_{\text{int}}(q,a)$ in (\ref{Hamiltonian}) can be one of such
observables.\footnote{More precisely, $\epsilon_i$ should be defined
by $x$ and $\pi$, but macroscopically the value of
$\epsilon_i(x,\pi)$ for a state $x$ comes from the time average for
$\pi (t)$ during $\mathit{\Delta} t$, and only the linearly
increasing component $a(t)$ in $\pi (t)$ is effective, as $\pi'(t)$
is a fast fluctuating time series.} Although the number and kind of
the observables $\epsilon_i$ alter depending on the
thermodynamic process concerned, we expect generally $\mathscr{N} \ll N$ for
the conventional thermodynamic system with Avogadro's number of
degrees of freedom.  Finally, more specifically, we write the
expression of $\tilde{\rho} (x,t)$ as
\begin{equation}
\tilde{\rho}(x,t) = \tilde{\rho}(\epsilon(x,a),t); \label{rho-bar
epsilon}
\end{equation}
in abbreviation, we will use the notation
$\tilde{\rho}(\epsilon(x,a),t)\equiv\tilde{\rho}(\epsilon,t)$.\footnote{
With the expression (\ref{rho-bar epsilon}) and (\ref{condition for
a0}), the condition (\ref{condition-time for rho bar}) can be
rewritten as
\begin{equation}
\tilde{\rho} (\epsilon(x,a+\mathit{\Delta} a),t + \mathit{\Delta} t)
\approx \tilde{\rho} (\epsilon(x,a),t)  + \sum_{i,j} \frac{\partial
\tilde{\rho} (\epsilon(x,a),t)} {\partial \epsilon_i} \frac{\partial
\epsilon_i} {\partial a_j}\mathit{\Delta} a_j + \frac{\partial
\tilde{\rho} (\epsilon(x,a),t)} {\partial t} \mathit{\Delta} t.
\end{equation}}
 Corresponding to the variation of the external parameter $a$, the
statistical properties of each dynamical variable $\epsilon_i$ are
expected to vary slowly.  Thus, we assume that we can control the
statistical properties of $\epsilon_i$ with an appropriate time series
$a(t)$ and the external thermal environment.

Using the cumulants \cite{Risken}, we can write
\begin{equation}
\tilde{\rho}(\epsilon(x,a),t) = \mspace{-1mu}\text{const}
\mspace{1mu} \cdot \int_{-\infty}^{\infty} \exp \bigg [ {-i k \cdot
x + \sum_{|m|=1}^{\infty}\frac{K_m}{m!} (i k)^m} \bigg ] dk ,
\end{equation}
where $k \equiv (k_1,k_2,\dots,k_{2N})$ and $k \cdot x \equiv
\sum_{i} k_i  x_i$, and  $m \equiv (m_1, m_2, \dots, m_{2N})$ is
multi-index; the cumulant $K_m$ is given as
\begin{equation}
K_m(a,t) = (-i)^{|m|} \left [ \frac{\partial^{|m|}}{{\partial k}^m}
\ln \sum_{|l|=0}^{\infty} \frac{\langle {x}^l \rangle}{l!} (i k)^l
\right ]_{k=0}
\end{equation}
with multi-index $l = (l_1, l_2, \dots, l_{2N})$ and the notation
$\langle {x}^l \rangle \equiv \int \tilde{\rho}(x,t)  {x}^l d{x}$.
Thus, we write $\tilde{\rho}(\epsilon,t)$ as a function of $K_m$:
$\tilde{\rho}(\epsilon,t)=\tilde{\rho}(\epsilon, K_m)$. Because
$K_m$ varies slowly as $a_i(t)$, we have
\begin{equation}\label{K_m condition}
K_m(a + \mathit{\Delta} a, t + \mathit{\Delta} t) \approx K_m(a, t)
+ \frac{dK_m(a, t)}{dt}  \mathit{\Delta} t.
\end{equation}

With multi-index $n=(n_1,n_2,\dots,n_{\mathscr{N}})$, let us define
$c_n$ as
\begin{equation}
\tilde{\rho}(x, t)=\tilde{\rho}(\epsilon, K_m)=
e^{\sum_{|n|=0}^{\infty}c_n  \epsilon^n}, \label{definition of c}
\end{equation}
i.e.,
\begin{equation}
c_n(a,t)= \left [ \frac{1}{n!}  \frac{\partial^{|n|} \ln
\tilde{\rho}(\epsilon, K_m)}{{\partial \epsilon}^n}\right
]_{\epsilon=0}.
\end{equation}
Then, considering (\ref{K_m condition}), $c_n$ varies slowly also as
\begin{equation}\label{condition for c_n}
\mathit{\Delta} c_n=c_n(a+\mathit{\Delta} a,t + \mathit{\Delta} t) -
c_n(a,t) \approx \frac{d c_n}{dt}  \mathit{\Delta} t.
\end{equation}
Finally we write $\tilde{\rho}(\epsilon,t)$ as a function of $c_n$:
\begin{equation}
\tilde{\rho}(\epsilon(x,a),t)=\tilde{\rho}(\epsilon(x,a), c_n).
\end{equation}

\section{Expression of entropy and
proof of the entropy theorem}\label{proof-of-entropy-theorem}
Now, after time $\mathit{\Delta} t$ from $t$, we have the ensemble
distribution $\tilde{\rho}(\epsilon(x,a+\mathit{\Delta} a), c_n
+\mathit{\Delta} c_n)$. Applying (\ref{condition for a0}),
(\ref{condition1 for rho bar}), (\ref{condition for c_n}), and the
normalization condition (\ref{norm for rho bar}), we have
{\setlength\arraycolsep{0pt}
\begin{eqnarray}
&&\int  (e^{\ln \tilde{\rho}(\epsilon(x,a+\mathit{\Delta} a), c_n
+\mathit{\Delta}
c_n)} -e^{\ln \tilde{\rho}(\epsilon(x,a), c_n)}  )dx \nonumber \\
= && \int \tilde{\rho}(\epsilon,c_n) \bigg ( \sum_{|n|=0}^{\infty} \frac{\partial
\ln \tilde{\rho}}{\partial c_n}  \mathit{\Delta} c_n + \sum_{i,j}
\frac{\partial \ln \tilde{\rho}}{\partial \epsilon_j} \frac{\partial
\epsilon_j}{\partial a_i} \mathit{\Delta}
a_i \bigg )dx \nonumber \\
= && 0,
\end{eqnarray}
}where $\tilde{\rho} \equiv \tilde{\rho}(\epsilon,c_n)$.  Using
(\ref{definition of c}) and $\sum_{j}\partial_{\epsilon_j} \ln
\tilde{\rho} \cdot
\partial_{a_i} \epsilon_j= \sum_{|n|=1}^{\infty}c_n  \partial_{a_i} \epsilon^n$,
and noting that, since the averaged quantities
\begin{equation}
\langle \epsilon^n \rangle \equiv \int \tilde{\rho}(x,t)
{\epsilon}^n d{x}
\end{equation}
vary slowly as (\ref{condition1 for rho bar}), we have $\langle
\epsilon^n \rangle  \mathit{\Delta} c_n= \mathit{\Delta} (\langle
\epsilon^n \rangle  c_n) - c_n  \mathit{\Delta} \langle \epsilon^n
\rangle$,  we obtain
\begin{equation}
\sum_{|n|=1}^{\infty}c_n   \mathit{\Delta} \langle \epsilon^n
\rangle - \sum_{|n|=1}^{\infty}  \sum_{i}  c_n  \frac{\partial
\langle \epsilon^n \rangle}{\partial a_i}   \mathit{\Delta} a_i
-\mathit{\Delta} \Big ( \sum_{|n|=0}^{\infty} c_n  \langle
\epsilon^n \rangle \Big )=0. \label{thermodynamic formula}
\end{equation}
Comparing with the approach for the quasi-equilibrium process given by
Gibbs and Einstein \cite{Gibbs,Einstein},
it can be understood that the first  and second term respectively correspond to the terms
describing the energy variation and the work performed on external
system in the equilibrium thermodynamics, and the third term
corresponds to the variation of the Gibbs entropy.

Let us denote the third term in (\ref{thermodynamic formula}) as
\begin{equation}
\bar{S}(t) \equiv -  \sum_{|n|=0}^{\infty} c_n  \langle \epsilon^n
\rangle = -\int  \tilde{\rho}(x,t) \ln \tilde{\rho}(x,t) dx.
\label{defintion of S-bar}
\end{equation}
Using the simplified notation $\int_{t}^{t+\mathit{\Delta} t} dt'
\int_{x}^{x+ \mathit{\Delta} x}  dx'\cdots \equiv
\int_{t,x,\dots}^{\mathit{\Delta}}dt' dx' \cdots $,\footnote{As a
special case of this notation, we will use $\int_{x}^{\mathit{\Delta}}dx'
  \equiv \int_{x}^{x+ \mathit{\Delta} x} dx'
$ and $\int_{t}^{\mathit{\Delta}}dt'   \equiv
\int_{t}^{t+ \mathit{\Delta} t} dt' $.} and by the
condition (\ref{condition1 for rho bar}) for the slowly varying
property of $\tilde{\rho}(x,t)$,  in zeroth order of $\mathit{\Delta} t$,
we can write \footnote{In obtaining the
second line of (\ref{calculation of entropy}), we have used
\begin{equation}\label{calculation formula1}
\int dx  \rho (x,t) \cdots = \int dx  \frac{1}{\mathit{\Delta}
\mathit{\Gamma}}\int_{x}^{\mathit{\Delta}} dx'  \rho (x',t)\cdots.
\end{equation}}
{\setlength\arraycolsep{1.4pt}
\begin{eqnarray}\label{calculation of entropy}
\displaystyle \bar{S}(t)& =& \displaystyle -\frac{1}{\mathit{\Delta}
t \mathit{\Delta}\mathit{\Gamma}} \int dx \int_{t,x}^{\mathit{\Delta}}dt' dx'
\rho(x',t') \ln \tilde{\rho}(x,t) \nonumber \\
\displaystyle &=& \displaystyle - \frac{1}{\mathit{\Delta} t
}\int_{t}^{\mathit{\Delta}} dt' \int dx  \rho (x,t') \ln
\frac{1}{\mathit{\Delta} t  \mathit{\Delta}
\mathit{\Gamma}}\int_{t, x}^{\mathit{\Delta}} ds  dy \rho(y,s).
\end{eqnarray}
}Here, from the Liouville equation (\ref{Liouville's equation}),
$\rho(y,s)$ in the last integral can be expressed as
{\setlength\arraycolsep{1.4pt}
\begin{eqnarray}
\rho (y,s) &=& \hat{T} \big [ e^{-i\int_{t}^{s}L(y, \tau)
d\tau} \big ] \rho (y,t) \nonumber \\
&=& \int dy'  \rho (y',t) \hat{T}  \big [
e^{-i\int_{t}^{s}L(y,\tau) d\tau} \big ] \delta (y-y').
\label{expression of rho}
\end{eqnarray}
}Thus, finally we obtain {\setlength\arraycolsep{0pt}
\begin{eqnarray}\label{macroscopic Gibbs entropy}
\bar{S}(t) = - \frac{1}{\mathit{\Delta} t}\int_{t}^{\mathit{\Delta}}
dt' &&\int dx  \rho (x,t') \ln   \int dy'
\rho (y',t) \nonumber \\
&& \times  \frac{1}{\mathit{\Delta} t  \mathit{\Delta}
\mathit{\Gamma}}\int_{t,x}^{\mathit{\Delta}} ds  dy \hat{T} \big [
e^{-i\int_{t}^{s}L(y,\tau) d\tau} \big ]  \delta (y-y').
\end{eqnarray}
}And correspondingly the microdynamical expression for entropy, $S$, should be given
as
\begin{equation}\label{Gibbs entropy}
S(t) =  - \int  dx  \rho (x,t)  \ln  \int dx'  \rho (x',t)
\frac{1}{\mathit{\Delta} t {\mathit{\Delta} \mathit{\Gamma}}}
\int_{t,x}^{\mathit{\Delta}} ds  dy \hat{T} \big [
e^{-i\int_{t}^{s}L(y,\tau) d\tau} \big ] \delta (y-x').
\end{equation}
In the above expression (\ref{Gibbs entropy}), it should be noted
that, considering the error range $-\frac{\mathit{\Delta} t}{2}
<\xi_0 < \frac{\mathit{\Delta} t}{2}$, the last integral term
expresses the probability density $p_{\mathit{\Delta} t}(x|x',t)$ to
observe (in macroscopic realm) that the system initially
positioned at $x'$ in $\mathit{\Gamma}$ at $t$ is found at position
$x$ after time elapse $\frac{\mathit{\Delta} t}{2} $, i.e.,
\begin{equation}
p_{\mathit{\Delta} t}(x|x',t)= \frac{1}{\mathit{\Delta} t
{\mathit{\Delta} \mathit{\Gamma}}} \int_{t,x}^{\mathit{\Delta}} ds
dy \hat{T} \big [  e^{-i\int_{t}^{s}L(y,\tau) d\tau}\big ]  \delta
(y-x') \label{expression of p}
\end{equation}
and it is the quantity describing the irreversible variation of the
dynamical system initially positioned at $x'$.\footnote{If we use
the more precise definition (\ref{definition2 of rho bar}) for
$\tilde{\rho}$, the expression (\ref{expression of p}) is replaced
by
\begin{equation}
p_{\mathit{\Delta} t}(x|x',t)=  \int_{- \infty}^{\infty} \int_{-
\infty}^{\infty} d \xi_0  d \xi  f(\xi, \xi_0) \hat{T} \Big [
e^{-i\int_{t-\frac{\mathit{\Delta} t}{2}}^{t+\xi_0}L(x+\xi,\tau)
d\tau}\Big ]  \delta (x + \xi - x').
\end{equation}}  Because of the fast
chaotic and ergodic motions generating the results (\ref{fastmotion
condition}) and (\ref{fastmotion condition2}), $p_{\mathit{\Delta}
t}(x|x',t)$ is different from $\delta$-function, as the range of $x$
over which $p_{\mathit{\Delta} t}(x|x',t) \not = 0$ cannot be
covered by $\mathit{\Delta} x$, and the irreversibility contained in
the Liouville operator $L(x,a,t)$ exactly corresponds to this
irreversible functional variation  from the initial distribution
$\delta (x-x')$. As the initial position $x'$ is changed,
$p_{\mathit{\Delta} t}(x|x',t)$ gives the complete information on
the irreversibility of the entire dynamics at $t$.   In the following
analysis, with the given above entropy expression, we will prove the
entropy theorem for the completely isolated system.

Let us consider the completely isolated system as $\pi_i =
\mspace{-1mu}\text{const}$. Then $p_{\mathit{\Delta} t}(x|x',t)$
is independent of $t$:
\begin{equation}
p_{\mathit{\Delta} t}(x|x')= \frac{1}{\mathit{\Delta} t
{\mathit{\Delta} \mathit{\Gamma}}} \int_{0,x}^{\mathit{\Delta}} ds
dy  e^{-i L(y) s}  \delta (y-x'). \label{closed expression of p}
\end{equation}
For the fast chaotic and ergodic motions in the completely isolated
system, we request that these complex motions make the following
statement established:
\begin{itemize}
\item[\textasteriskcentered] $p_{\mathit{\Delta} t}(x|x')$ is a slowly
varying function for $x$ and $x'$ as $\tilde{\rho}$ in the sense of
(\ref{condition1 for rho bar}).\footnote{Finally, the
   condition for the fast chaotic and ergodic dynamics of Hamiltonian
   $\mathscr{H}$ to prove the entropy theorem can be summarized
   as follows.  As the chaotic and ergodic motions generate the presupposed
   results (\ref{fastmotion condition}) and (\ref{fastmotion condition2})
   and give the property (\ref{condition1 for rho bar})
   to $\tilde{\rho} (x,t)$, by these complex motions,
   \begin{itemize}\label{condition_for_proof}
   \item[1] $p_{\mathit{\Delta} t}(x|x')$ cannot be a $\delta$-function
   in macroscopic sense: i.e., for each point $x$ on the
   appropriate energy surface in $\mathit{\Gamma}$ (corresponding
   to the initial conditions of the completely isolated dynamical system),
   there exists some macroscopically distinguishable different point
   $x'$ from $x$ in spite of the measurement errors such that
   $p_{\mathit{\Delta} t}(x|x') \not = 0$. (In our case, $x'$ should
   satisfy $|x_i-x'_i| > \mathit{\Delta} x_i$ for $i=1,2,\dots,2N$.)
   \item[2] $p_{\mathit{\Delta} t}(x|x')$ is a slowly varying function
    for $x$ and $x'$ as $\tilde{\rho}$ in the sense of
    (\ref{condition1 for rho bar}).
   \end{itemize}}
\end{itemize}
Then, by the above  slowly varying condition for $p_{\mathit{\Delta}
t}(x|x')$ and the identity
\begin{equation}\label{L relation for delta}
e^{-i L(y) s}  \delta (y-x)=e^{i L(x) s}  \delta (y-x),
\end{equation}
we can write as follows: {\setlength\arraycolsep{1.4pt}
\begin{eqnarray}\label{inverse of p}
p_{\mathit{\Delta} t}(x|x') &=& \frac{1}{\mathit{\Delta} t
{\mathit{\Delta} \mathit{\Gamma}}} \int_{x,0}^{\mathit{\Delta}}dy ds
e^{-i L(y) s}  \delta (y-x') \nonumber \\
&=&\frac{1}{\mathit{\Delta} t  {\mathit{\Delta} \mathit{\Gamma}}^2 }
\int_{x',x,0}^{\mathit{\Delta}}dy'  dy ds  e^{-i L(y) s}  \delta
(y-y')
\nonumber \\
&=&\frac{1}{\mathit{\Delta} t  {\mathit{\Delta} \mathit{\Gamma}}^2 }
\int_{x,x',0}^{\mathit{\Delta}}dy  dy' ds  e^{i L(y') s}  \delta
(y-y')
\nonumber \\
&=& \frac{1}{\mathit{\Delta} t  {\mathit{\Delta} \mathit{\Gamma}}^2}
\int_{x,x',0}^{\mathit{\Delta}}dy  dy' ds e^{-i L(y') (s
- \mathit{\Delta} t)}  \delta (y-y')
\nonumber \\
&=& \frac{1}{\mathit{\Delta} t  {\mathit{\Delta} \mathit{\Gamma}}^2}
\int_{x,x',0}^{\mathit{\Delta}}dy  dy' ds e^{-i L(y') s -i L(y)
\mathit{\Delta} t} \delta (y-y')
\nonumber \\
&=& \frac{1}{\mathit{\Delta} \mathit{\Gamma}}
\int_{x}^{\mathit{\Delta}}dy  e^{-i L(y) \mathit{\Delta} t}
p_{\mathit{\Delta} t}(x'|y).
\end{eqnarray}
}On the other hand, from (\ref{definition of rho bar}) and
(\ref{expression of rho}), $\tilde{\rho}(x,t)$ can be expressed as
\begin{equation}\label{tilde rho by p}
\tilde{\rho}(x,t) = \int dx'  \rho (x',t)  p_{\mathit{\Delta}
t}(x|x'),
\end{equation}
from which, using  $\rho(x',t- \mathit{\Delta} t)=e^{i L(x')
\mathit{\Delta} t} \rho (x',t)$ and the Hermitian property of the
operator $L(x')$, we obtain {\setlength\arraycolsep{1.4pt}
\begin{eqnarray}\label{condition1 for p}
\tilde{\rho}(x,t-\mathit{\Delta} t) - \tilde{\rho}(x,t)= \int dx'
\rho (x',t)  (e^{-i L(x') \mathit{\Delta} t}-1) p_{\mathit{\Delta}
t}(x|x')
\end{eqnarray}
}and similarly, using the Liouville equation (\ref{Liouville's
equation}), {\setlength\arraycolsep{1.4pt}
\begin{eqnarray}\label{condition2 for p}
\frac{\partial \tilde{\rho}(x,t)}{\partial t} = \int dx'  \rho
(x',t) i  L(x')  p_{\mathit{\Delta} t}(x|x').
\end{eqnarray}
}With $\delta$-function $\delta(x) \equiv \delta(x_1)
\delta(x_2)\cdots\delta(x_{2N})$, let us define the macroscopic
$\delta$-function $\tilde{\delta}(x)$ as
\begin{equation}\label{macroscopic delta function}
\tilde{\delta} (x) \equiv \frac{1}{\mathit{\Delta}
\mathit{\Gamma}}\int_{x}^{\mathit{\Delta}}  \delta (\mathit{\Delta} x
- y)  dy.
\end{equation}
Then, substituting  (\ref{condition1 for p}) and (\ref{condition2
for p}) into condition (\ref{condition1 for rho bar}) and replacing
$\rho(x',t)$ by $\tilde{\delta}(x'-y')$ with the notion that
(\ref{condition1 for p}) and (\ref{condition2 for p}) hold for
arbitrary $\rho(x',t)$, we obtain
\begin{equation}\label{condition3 for p}
\frac{1}{\mathit{\Delta} \mathit{\Gamma}}
\int_{y'}^{\mathit{\Delta}}dx'  e^{-i L(x') \mathit{\Delta} t}
p_{\mathit{\Delta} t}(x|x') \approx p_{\mathit{\Delta}
t}(x|y')- \mathit{\Delta} t \cdot \frac{i}{\mathit{\Delta}
\mathit{\Gamma}} \int_{y'}^{\mathit{\Delta}}dx'  L(x')
p_{\mathit{\Delta} t}(x|x')
\end{equation}
for the description of
$\tilde{\rho}(x,t)$ as conditioned by (\ref{condition1 for rho bar}).
Finally, by (\ref{inverse of p}) and the approximation (\ref{condition3
for p}), in macroscopic realm, we establish

\begin{equation}\label{condition for p}
p_{\mathit{\Delta} t}(x|x') = p_{\mathit{\Delta} t}(x'|x).
\end{equation}

Also, since $\tilde{\rho}(x,t)$ is expressed as in (\ref{tilde rho
by p}) for the completely isolated system, integrating both sides of
(\ref{tilde rho by p}) by $\frac{1}{\mathit{\Delta} t}
\int_{t}^{\mathit{\Delta}}dt'$ and using the relation given from
condition (\ref{condition1 for rho bar})
{\setlength\arraycolsep{1.4pt}
\begin{eqnarray}\label{integration-condtion-rho-tilde}
\frac{1}{\mathit{\Delta} t}\int_{t}^{\mathit{\Delta}} dt'
\tilde{\rho}(x,t') &\approx& \frac{1}{\mathit{\Delta}
t}\int_{t}^{\mathit{\Delta}} dt' \left [ \tilde{\rho}(x,t) +
\partial_{t}\tilde{\rho}(x,t) \cdot (t'-t) \right ] \nonumber \\
&\approx& \tilde{\rho}(x,t + { \textstyle\frac{\mathit{\Delta}
t}{2}})
\end{eqnarray}
}and formula (\ref{calculation formula1}), we can write
{\setlength\arraycolsep{1.4pt}
\begin{eqnarray}\label{tilde rho expressed by p(x|x)}
\tilde{\rho}(x,t + {\textstyle \frac{\mathit{\Delta} t}{2}})& = &
\frac{1}{\mathit{\Delta} t} \int_{t}^{\mathit{\Delta}}dt'
\int dx'  \rho (x',t')  p_{\mathit{\Delta} t}(x|x') \nonumber \\
& = & \int dx'  \frac{1}{\mathit{\Delta} t  \mathit{\Delta}
\mathit{\Gamma}} \int_{t,x'}^{\mathit{\Delta}}dt'  dy
\rho (y,t')  p_{\mathit{\Delta} t}(x|y) \nonumber \\
& = & \int dx'  \tilde{\rho}(x',t)  p_{\mathit{\Delta} t}(x|x'),
\end{eqnarray}
}because $p_{\mathit{\Delta} t}(x|x')$ is a slowly varying function
for $x$ and $x'$ as $\tilde{\rho}$.
%
%
Then, with (\ref{tilde rho expressed by p(x|x)}) and
 (\ref{condition1 for rho bar}), we arrive the result
\begin{equation}\label{equation of tilde rho}
\frac{\partial \tilde{\rho}(x,t)}{\partial t} =
\frac{2}{\mathit{\Delta} t} \int dx' p_{\mathit{\Delta} t}(x|x')
\left [ \tilde{\rho}(x',t) -\tilde{\rho}(x,t)\right ],
\end{equation}
where we have used $\int dx' p_{\mathit{\Delta} t}(x|x')=\int dx'
p_{\mathit{\Delta} t}(x'|x)=1$.

 Thus, from the definition
(\ref{defintion of S-bar}) and (\ref{equation of tilde rho}),
\begin{equation}
\frac{ d \bar{S}(t)}{d t} = -\frac{2}{\mathit{\Delta} t} \int  dx
\int dx'   p_{\mathit{\Delta} t}(x|x')\left [\tilde{\rho}(x',t)
-\tilde{\rho}(x,t)\right ] \left [ 1 + \ln {\tilde{\rho}}(x,t)\right
].
\end{equation}
Exchanging $x$ and $x'$, we obtain a different expression of $\frac{
d \bar{S}(t)}{d t}$, and, adding these two expressions and using
(\ref{condition for p}),
\begin{equation}
 \frac{ d \bar{S}(t)}{d t} = -\frac{1}{\mathit{\Delta} t}  \int dx \int
dx' p_{\mathit{\Delta} t}(x|x') \left [ \tilde{\rho}(x',t)
-\tilde{\rho}(x,t)\right ] \ln
{\frac{\tilde{\rho}(x,t)}{\tilde{\rho}(x',t)}}.
\end{equation}
Because, in the sense of statement 1 of footnote \ref{condition_for_proof},
 $p_{\mathit{\Delta} t}(x|x')$ given by (\ref{closed
expression of p}) cannot be a $\delta$-function in macroscopic realm
as is defined by (\ref{macroscopic delta function}),
and the
equilibrium state of the completely isolated system is identified as
$\tilde{\rho}(x,t)=\mspace{-1mu}\text{const}$ on an appropriate
energy surface, we have
\begin{equation}
\frac{ d \bar{S}(t)}{d t} > 0, \label{entropy theorem}
\end{equation}
if the system is in a non-equilibrium state as $\tilde{\rho}(x,t) \not
=\mspace{-1mu}\text{const}$ on the energy surface; for the
equilibrium state, $\frac{ d \bar{S}(t)}{d t} = 0$.

Consequently, as is manifested in comparison with the process of
derivation of the Boltzmann transport equation
\cite{Huang,Boltzmann},  the term $p_{\mathit{\Delta} t}(x|x')$
takes the role corresponding to the scattering matrix in the
Boltzmann transport equation.
%
%
%
%
Similarly as the scattering matrix in Boltzmann's theory describes
the effect of the local intermolecular interaction given by a physical
potential (such as Coulomb potential), the term $p_{\mathit{\Delta}
t}(x|x')$ describes the effect of the fast chaotic and ergodic
motions generating (\ref{fastmotion condition}) and (\ref{fastmotion
condition2}).  If we consider the systems initially positioned in
the volume $(x,x + \mathit{\Delta} x)$ in $\mathit{\Gamma}$, while
these initial conditions are macroscopically un-discriminable as the
differences between them fall within the error range, the fast
complex motions change the each system with the finely
differentiated initial conditions to be respectively in completely
different positions in $\mathit{\Gamma}$  after time elapse
$\mathit{\Delta} t$ (or $\frac{\mathit{\Delta} t}{2}$). And
macroscopically this dynamical process related with the fast complex
motions can be understood as an $\frac{N}{3}$-body scattering-like
process.  To gain a more physically intuitive picture for the
calculation process of the entropy theorem, we rewrite entropy
$\bar{S}$ given by (\ref{defintion of S-bar}), (\ref{macroscopic
Gibbs entropy}) and (\ref{closed expression of p}) as follows using
the relation (\ref{tilde rho expressed by p(x|x)}) in zeroth order
of $\mathit{\Delta} t$:
\begin{equation}\label{entropy3 by ensemble potential}
\bar{S}(t)=-\frac{1}{2}\int dx\int dx'  \tilde{\rho}(x,t)
V^{\mathit{\Gamma}}(x,x',t) \tilde{\rho} (x',t),
\end{equation}
where
\begin{equation}\label{aymmetric solution}
V^{\mathit{\Gamma}}(x,x',t)= p_{\mathit{\Delta} t}(x|x') \left [
\frac{\ln \tilde{\rho} (x,t)}{\tilde{\rho} (x,t)} +  \frac{\ln
\tilde{\rho} (x',t)}{\tilde{\rho} (x',t)}\right ].
\end{equation}
If we interpret $V^{\mathit{\Gamma}}(x,x',t)$ as a potential
interacting between two ensemble elements respectively positioned at
$x$ and $x'$ in $\mathit{\Gamma}$ at time $t$, the entropy theorem
(\ref{entropy theorem}) simply means that the ensemble system always
behaves to decrease the total potential energy induced by the
ensemble potential $V^{\mathit{\Gamma}}(x,x',t)$. And, since
$p_{\mathit{\Delta} t}(x|x') \not\approx \tilde{\delta}(x-x')$ by
the fast chaotic and ergodic motions generating (\ref{fastmotion
condition}) and (\ref{fastmotion condition2}), it means that
generally the elements of the ensemble interact with each other
through the potential $V^{\mathit{\Gamma}}(x,x',t)$; since we can
expect from definition (\ref{closed expression of p}) that
$p_{\mathit{\Delta} t}(x|x')$ and $V^{\mathit{\Gamma}}(x,x',t)$ have
the tendency to decrease as the distance $|x-x'|$ increases in
$\mathit{\Gamma}$, we can say that the interaction by
$V^{\mathit{\Gamma}}(x,x',t)$ is overall repulsive.\footnote{On the
contrary, in the original Gibbs entropy expression,
the ensemble elements in $\mathit{\Gamma}$ are treated as they are completely
independent of one another as the original expression is written in the form
$-\int dx \rho (x,t) \ln { \int dx'\rho (x',t)} \delta (x-x')$, that
is, $p_{\mathit{\Delta} t}(x|x') = \delta(x-x')$.}
In macroscopic observation realm, the fast chaotic and ergodic
motions generating the results (\ref{fastmotion condition}) and
(\ref{fastmotion condition2}) emerges as a potential giving the
interaction between the ensemble elements in phase space
$\mathit{\Gamma}$; In particular, if the ensemble system is in equilibrium
($\tilde{\rho}=\mspace{-1mu}\text{const}$ on an appropriate energy
surface), the potential is given by $V^{\mathit{\Gamma}}(x,x',t)
\propto p_{\mathit{\Delta} t}(x|x')$.   We use the notation
$\phi_{\mathit{\Delta} t}(x,x',t)$ instead of $p_{\mathit{\Delta}
t}(x|x',t)$ to emphasize the meaning of $p_{\mathit{\Delta}
t}(x|x',t)$ as an ensemble potential originated from dynamic
correlation, and the microscopic expression of entropy is written as
\begin{equation}\label{Gibbs entropy2}
S(t) = -\int dx  \rho (x,t)  \ln {  \int dx'  \rho (x',t)
\phi_{\mathit{\Delta} t}(x,x',t)},
\end{equation}
where $\phi_{\mathit{\Delta} t}(x,x',t)\equiv p_{\mathit{\Delta}
t}(x|x',t)$ is given by the expression (\ref{expression of p}).

Finally, we point out that it is logically natural that the
expression of entropy contains the parameter $\mathit{\Delta} t$
which specifies the scale of the given measurement system, in that
the entropy increment of a system is essentially related with the
microdynamical components which cannot be described in a deterministic
way under the given measurement system: this fine dynamical
information deterministically un-catchable by the given measurement
system increases as the measuring scale $\mathit{\Delta} t$
increases.  Under the given measurement system, the observer loses
such fine dynamical information, and, \emph{only when there is the loss of
the deterministic dynamical information by measurement imperfection,
the entropy increment arises}. For the perfect observer
($\mathit{\Delta} t=0$ and $\mathit{\Delta} x =0$), the entropy expression
becomes exactly the Gibbs entropy, i.e., $\bar{S}(t)=-\int dx  \rho
(x,t) \ln {\rho (x,t)}$, and thus there is no increment of entropy:
$\frac{d \bar{S}}{dt}=0$
as $\bar{S}$ is dynamically invariant.

\section{Meaning of the entropy theorem in phase space
dynamics}\label{meaning-of-entropy-theorem}
From the Liouville equation (\ref{Liouville's equation}), we can
obtain the description for the macroscopic ensemble distribution
$\tilde{\rho} (x,t)$ in $\mathit{\Gamma}$. Integrating both sides of
the Liouville equation (\ref{Liouville's equation}) by
$\int_{t,x}^{\mathit{\Delta}} dt' dx'$ and noting the expression
(\ref{expression of rho}), we have the results as
%
%
\begin{equation}
\frac{\partial \tilde{\rho} (x,t)}{\partial t}= -\sum_{i}
\frac{\partial}{\partial x_i}  \tilde{\rho}(x,t) \tilde{v}_i(x,t),
\label{continuity equation for rho-bar}
\end{equation}
where, for the completely isolated system, $\tilde{v}_i(x,t)$ is
given as {\setlength\arraycolsep{1.4pt}
\begin{eqnarray}\label{macroscopic velocity1}
\tilde{v}_i(x,t) &\equiv&  \frac{1}{\tilde{\rho} (x,t)
\mathit{\Delta} t  \mathit{\Delta} \mathit{\Gamma}}
\int_{t,x}^{\mathit{\Delta}}dt' dx'  \dot{x}_i(x') \rho(x',t')
\nonumber \\
&=& \int dx'  \tilde{v}_i^{o}(x,x')  \frac{\rho (x',t)}{\tilde{\rho}
(x,t)}
\end{eqnarray}
}with the definition
\begin{equation}\label{macroscopic velocity0}
\tilde{v}_i^{o}(x,x') \equiv \frac{1}{\mathit{\Delta} t
{\mathit{\Delta} \mathit{\Gamma}}} \int_{0,x}^{\mathit{\Delta}} ds
dy  \dot{x}_i(y)  e^{-i L(y) s}  \delta (y-x'),
\end{equation}
where $\dot{x}(x) \equiv
(\dot{x}_1(x),\dot{x}_2(x),\dots,\dot{x}_{2N}(x)) \equiv
(\dot{q}(x),\dot{p}(x))$ denotes the velocity field in
$\mathit{\Gamma}$ given by Hamilton's equations of motion for
$\mathscr{H}(x)$, i.e.,
\begin{equation}
\dot{x}(x) = { \left ( \partial_{p_1}
\mathscr{H},\dots,\partial_{p_N} \mathscr{H},-\partial_{q_1}
\mathscr{H},\dots,-\partial_{q_N} \mathscr{H} \right )};
\end{equation}
if we integrate both sides of (\ref{macroscopic velocity1})  by $\int_{t}^{\mathit{\Delta}}dt'$ and
note that $\tilde{v}_i^{o}$ and $\tilde{v}_i$ are slowly varying
functions as $\tilde{\rho} (x,t)$,  then,  with formula (\ref{calculation formula1}) and
a similar process to that in (\ref{integration-condtion-rho-tilde})
and (\ref{tilde rho expressed by p(x|x)}), in zeroth order of  $\mathit{\Delta} t$,
(\ref{macroscopic velocity1}) can be rewritten  as \footnote{As is indicated by the
expression (\ref{macroscopic velocity}), $\mathit{\Delta}
\mathit{\Gamma} \cdot \tilde{v}_i^{o}(y,y)$ is the velocity field
(at position $y$ in $\mathit{\Gamma}$) observed for the ensemble of
the systems which are prepared to be in state $y$ under the
macroscopic measurement, i.e., $\tilde{\rho} (x) = \tilde{\delta}
(x-y)$.  And, for the ensemble system prepared as $\tilde{\rho} (x)
= \frac{1}{2}  \tilde{\delta} (x-y) + \frac{1}{2}
\tilde{\delta}(x-y')$, $\mathit{\Delta} \mathit{\Gamma} \cdot
\tilde{v}_i^{o}(y,y')$ is the contribution to the velocity field at
$y$ by the ensemble elements positioned at $y'$.}
\begin{equation}\label{macroscopic velocity}
\tilde{v}_i(x,t) = \int dx'
\tilde{v}_i^{o}(x,x')  \frac{\tilde{\rho} (x',t)}{\tilde{\rho}
(x,t)}.
\end{equation}

That is, $\tilde{v}_i(x,t)$ is the velocity field (in
$\mathit{\Gamma}$) observed in macroscopic point of view, and the
averaged macroscopic motion in $\mathit{\Gamma}$ is given as
\begin{equation}\label{closed macroscopic equation of motion}
\frac{dx_i}{dt} = \tilde{v}_i(x,t).
\end{equation}
Then, by the definition of entropy (\ref{defintion of S-bar}) and
(\ref{continuity equation for rho-bar}), we have
\begin{equation}\label{time derivative of entropy}
\frac{ d \bar{S}(t)}{d t} = \int \tilde{\rho}(x,t) \sum_{i}
\frac{\partial \tilde{v}_i(x,t)}{\partial x_i}  dx.
\end{equation}
Thus, (\ref{entropy theorem}) means that, for the completely
isolated system,  the macroscopically observed motion of the
ensemble in $\mathit{\Gamma}$ is not incompressible any more:
generally,
\begin{equation}
\sum_{i} \frac{\partial \tilde{v}_i(x,t)}{\partial x_i} \not = 0,
\end{equation}
and, in macroscopic realm, the local volume in $\mathit{\Gamma}$
averagely expands, if the system is in a non-equilibrium state
($\tilde{\rho} \not =\mspace{-1mu}\text{const}$ on an energy
surface). Finally, if we treat the combined system of $\mathscr{H}$
and its environment system as a closed (i.e., completely isolated)
Hamiltonian
system, the entropy theorem can be restated as \emph{averagely the
local volume in the total phase space always expands, if the total
system is in a non-equilibrium state as the ensemble distribution is
not constant on an appropriate energy surface}.

\section{Thermodynamic relation}
Recalling the assumption that we can macroscopically control the
statistical properties of $\epsilon_i$ in (\ref{rho-bar epsilon}),
let us consider the process where $a(t)$ and the external thermal
environment are given such that they induce only the change of energy
value $E \equiv H(x) + V_{\text{int}}(q,a)$ of the system
(\ref{Hamiltonian}): for
$\tilde{\rho}(E,\epsilon_2,\epsilon_3,\dots,t)$, we consider
the variation of $E$ only as $\epsilon_i = \mspace{-1mu}\text{const}$.
Then, the result corresponding to (\ref{thermodynamic formula}) is
simply obtained by replacing $\epsilon$ in (\ref{thermodynamic
formula}) with $E$:
\begin{equation}
\mathit{\Delta} \bar{S} = \sum_{n=1}^{\infty} \frac{1}{T_n}
\mathit{\Delta} \bar{E}^{(n)} + \sum_{n=1}^{\infty} \frac{1}{T_n}
\sum_{i} \bar{F}^{(n)}_{i}  \mathit{\Delta} a_i,
\label{thermodynamic formula2}
\end{equation}
where we have used the notations $\bar{E}^{(n)} \equiv \langle
E^n \rangle $, $T_n \equiv -\frac{1}{c_n}$, and
\begin{equation}
\bar{F}^{(n)}_{i} \equiv - \frac{\partial \langle E^n
\rangle}{\partial a_i}. \label{definition of F^(n)}
\end{equation}
In non-equilibrium processes, we have (\ref{thermodynamic formula2})
as a relation directly corresponding to the thermodynamic relation
in the equilibrium thermodynamics.

The formulation (\ref{thermodynamic formula2}) means that, in
general non-equilibrium thermodynamic processes, in addition to the
average energy $\bar{E}^{(1)}$, we should treat each statistical
property $\bar{E}^{(n)}\equiv\langle E^n \rangle $
of $E$ ($n=2,3,\dots$) as
an \emph{independent} macroscopic thermodynamic variable.  Also, to
describe the mechanical interaction with the external environment,
we should treat the new $\bar{E}^{(n)}$-corresponding forces
$\bar{F}^{(n)}_{i}$ defined by (\ref{definition of F^(n)}) in
addition to the conventional force $\bar{F}^{(1)}_{i}= -
\frac{\partial \langle E \rangle}{\partial a_i} $. In this case, the
thermal characteristic of the system is completely determined by the
quantities $T_n$, which are given by the relation (\ref{thermodynamic
formula2}) as
\begin{equation}
\frac{1}{T_n} =  \frac{\partial \bar{S}}{\partial \bar{E}^{(n)}},\\
\end{equation}
where $n = 1,2,\dots$ ($T_1$ is the temperature in the equilibrium
thermodynamics). Keeping the conventional understanding of
temperature as a quantity to describe the thermal characteristic of
a system in relation with its entropy and energy variation,
(\ref{thermodynamic formula2}) means that, in general
non-equilibrium processes, the thermal characteristic of a system is
specified by many temperatures $T_n$. That is, generally the
temperature of a non-equilibrium thermodynamic system should be
defined not by a single scalar quantity but by an array of
temperatures
\begin{equation}
\mathbf{T} \equiv (T_1, T_2, \dots, T_n, \dots),
\end{equation}
where each component $T_n$ is the measure of how the entropy
$\bar{S}$ increases as the statistical property $ \langle E^{n}
\rangle$ increases.
%

Let $\mathit{\Sigma}$ be the space defined by the ordered-tuple
$(\mathbf{T},a)$. Since the ensemble distribution
$\tilde{\rho}(x,t)$ at an instant is completely determined by the
ordered-tuple $(\mathbf{T},a)$, an arbitrary thermodynamic process for
a given system can be described as a path in the space
$\mathit{\Sigma}$. If a system interacting with a thermal reservoir
transforms from equilibrium state $(T_1,a)$ to equilibrium state
$(T'_1,a')$,\footnote{The notation $(T_1,a)$ is the abbreviation of
$(\mathbf{T},a)$
for the case that $\frac{1}{T_n}=0$ for $n \geq 2$ in $(\mathbf{T},a)$.}
the entropy increment of the system during the process is
path-independent in the space $\mathit{\Sigma}$, because entropy
$\bar{S}$ defined by (\ref{defintion of S-bar}) is a function of
$(\mathbf{T},a)$ as $\langle \epsilon^n \rangle$ is a function of
$(\mathbf{T},a)$; if $R$ and $I$ respectively represents a
reversible and an irreversible path connecting state $(T_1,a)$ to state
$(T'_1,a')$ in $\mathit{\Sigma}$, the entropy increment for the
reversible path $\int_{R}d\bar{S}$  is the same as that  for the
irreversible path $\int_{I}d\bar{S}$. Thus, by (\ref{thermodynamic
formula2}), we have
\begin{equation}
\int_{R} \frac{d Q^{(1)}}{T_1} = \int_{I} \frac{d Q^{(1)}}{T_1}
+\sum_{n=2}^{\infty} \int_{I} \frac{d Q^{(n)}}{T_n},
\label{equilibrium-nonequilibrium relation}
\end{equation}
where
\begin{equation}
d Q^{(n)} \equiv d \bar{E}^{(n)} + \sum_{i} \bar{F}^{(n)}_{i}  d
a_i.
\end{equation}
Let the thermal reservoir be quasi-ideal in the sense that the
ensemble PDF of the closed total system (containing the thermal
reservoir) is always given by the form
\begin{equation}\label{condition for thermal reservoir}
\tilde{\rho}_{\text{tot}} = e^{c_0 -
\frac{\mathcal{E}}{\mathcal{T}}-\sum_{n=1}^{\infty}
\frac{E^n}{T_n}},
\end{equation}
where $\mathcal{E}$ and $\mathcal{T}$ are respectively the energy
and (conventional equilibrium) temperature of the reservoir: the
reservoir always varies through a quasi-equilibrium process. In such
a case, for all irreversible processes where $d Q^{(1)}=0$, the entropy
variation of the closed total system is given as
\begin{equation}
\delta \bar{S}_{\text{tot}} = \sum_{n=2}^{\infty} \int_{I} \frac{d
Q^{(n)}}{T_n}> 0,
\end{equation}
where the inequality comes from (\ref{entropy theorem}). Thus, we
expect $\sum_{n=2}^{\infty} \int_{I} \frac{d Q^{(n)}}{T_n} > 0$ in
(\ref{equilibrium-nonequilibrium relation}), and  obtain the
classical relation
\begin{equation}
\int_{R} \frac{d Q^{(1)}}{T_1} > \int_{I} \frac{d Q^{(1)}}{T_1}
\end{equation}
for the case that the thermal reservoir is quasi-ideal as is
described in (\ref{condition for thermal reservoir}).

\section{Derivation of fluctuation theorem}

As we have concluded in section \ref{proof-of-entropy-theorem}
and \ref{meaning-of-entropy-theorem},
despite the fact that the microscopic Hamiltonian dynamics
enforces the phase space volume conservation as the Liouville theorem declares,
by the loss of the deterministic dynamical information by measurement imperfection,
the physical phenomena eventually emerge as this conservation is broken,
and the entropy increment can be understood as the total result of
the phase space volume expansion or contraction following each
trajectory in phase space.  In connection with the study of the fluctuation theorem \cite{Evans},
this understanding gives us a perspective to take more strict comprehension for the general character
of the theorem.
In this section,
the reconstruction or reinterpretation
(in the view point of the definite Hamiltonian dynamical microscopic description)
for the fluctuation theorem pursued in the reference \cite{Evans}
will be provided, based on the performed analysis in section
\ref{proof-of-entropy-theorem} and \ref{meaning-of-entropy-theorem}.

%
%
%

In the following argument, to guarantee the precise microscopic dynamical
time reversibility, we treat the combined system of the dynamical
system $H$ in (\ref{Hamiltonian}) and its external environment
system as a completely isolated Hamiltonian dynamical system, and
write Hamiltonian $\mathscr{H}$ as
\begin{equation}\label{total Hamiltonian}
\mathscr{H}(x)= H + V_{\text{int}} + H_{\text{ext}},
\end{equation}
where $H_{\text{ext}}$ denotes the Hamiltonian of the external
environment and $V_{\text{int}}$ is the interaction potential
between the system $H$ and $H_{\text{ext}}$ as previously
introduced.  The phase space $\mathit{\Gamma}$ represents the total
phase space defined by the canonical variables of $H$ and
$H_{\text{ext}}$, and $x \equiv (q_1, \dots, q_N, p_1, \dots,
p_N)$ represents a point in this total phase space $\mathit{\Gamma}$.

Then, for the completely isolated system $\mathscr{H}(x)$,  the
macroscopically observed dynamics in $\mathit{\Gamma}$ is given by
the equations of motion (\ref{closed macroscopic equation of
motion}) with the velocity field expressed as (\ref{macroscopic
velocity}); as previously argued, generally this velocity field is
not incompressible.  By the continuity equation (\ref{continuity
equation for rho-bar}) and (\ref{closed macroscopic equation of
motion}), we obtain
\begin{equation}\label{rho-tilde equation}
\frac{d \tilde{\rho}(x,t)}{d t} =  - \tilde{\rho}(x,t) \sum_{i}
\frac{\partial \tilde{v}_i(x,t)}{\partial x_i}.
\end{equation}
Let us consider an arbitrary trajectory $x(t')$ satisfying the
equations of motion (\ref{closed macroscopic equation of motion});
let $x(t')$ satisfy $x(t_0)=x_0$ and $x(t)=x$.  For the trajectory
$x(t')$, from (\ref{rho-tilde equation}), we have
\begin{equation}\label{rho-tilde expression}
\tilde{\rho}(x,t) = \tilde{\rho}(x_0,t_0)  e^{-
\int_{t_0}^{t}\sum_{i} \frac{\partial \tilde{v}_i(t')}{\partial x_i}
dt'},
\end{equation}
where $\frac{\partial \tilde{v}_i(t')}{\partial x_i} \equiv
\frac{\partial \tilde{v}_i(x(t'),t')}{\partial x_i}$.   In the above
equation, from the expression (\ref{time derivative of entropy}) for
$\frac{d \bar{S}}{dt}$, we can interpret the quantity $\sum_{i}
\frac{\partial \tilde{v}_i (x,t)}{\partial x_i}$ as the entropy
increment rate for a state $x$ at $t$, and
\begin{equation}\label{entropy increment}
\delta s  \equiv \int_{t_0}^{t}\sum_{i} \frac{\partial
\tilde{v}_i(t')}{\partial x_i} dt'
\end{equation}
represents the entropy increment for the trajectory $x(t')$ during
the time interval $[t_0,t]$.

For $x(t') \equiv (q(t'),p(t'))$, let us define
\begin{equation}
x^{*}(t') \equiv (q(t'),-p(t')).
\end{equation}
Then, the time-reversed time series of $x(t')$ is given as
\begin{equation}
x^{\dag}(t') \equiv x^{*}(t+t_0-t').
\end{equation}
During the time interval $[t_0,t]$, the probability density to
observe the time series $x(t')$ is
\begin{equation}
p(x(t'); [t_0,t])= \tilde{\rho}(x_0,t_0).
\end{equation}
On the other hand, \emph{if the equations of motion (\ref{closed
macroscopic equation of motion}) allow the time-reversed solution
of the solution $x(t')$ during $[t_0,t]$,} i.e., if $x^{\dag}(t')$
can be a solution of (\ref{closed macroscopic equation of motion})
during $[t_0,t]$, the probability density to observe the
time-reversed series $x^{\dag}(t')$ during $[t_0,t]$ is
\begin{equation}
p(x^{\dag}(t'); [t_0,t])= \tilde{\rho}(x^{*}(t),t_0).
\end{equation}
Thus, using (\ref{rho-tilde expression}) and (\ref{entropy
increment}), we have
\begin{equation}\label{fluctuation theorem1}
\frac{p(x^{\dag}(t'); [t_0,t])}{p(x(t'); [t_0,t])} =
\frac{\tilde{\rho}(x^{*}(t),t_0)}{\tilde{\rho}(x(t),t)} e^{-\delta
s}.
\end{equation}
In particular, if $x(t')$ starts and ends in the same stationary state as
$\tilde{\rho}(y,t_0)=\tilde{\rho}(y,t)=\tilde{\rho}(y)$,
(\ref{fluctuation theorem1}) becomes
\begin{equation}\label{fluctuation theorem2}
\frac{p(x^{\dag}(t'); [t_0,t])}{p(x(t'); [t_0,t])} =
\frac{\tilde{\rho}(x^{*})}{\tilde{\rho}(x)}  e^{-\delta s},
\end{equation}
where generally the stationary state $\tilde{\rho}(y)$ can be
different from the equilibrium state expressed by
$\tilde{\rho}(y)=\mspace{-1mu}\text{const}$ on an energy
surface.\footnote{ For example, if the external environment system
is constituted by two ideal thermal reservoirs with different
temperatures from each other, where `ideal' means that they have
infinite energy capacity, then the time length necessary to
equilibrate the system $\mathscr{H}(x)$ becomes infinite, and
$\tilde{\rho}(y)$ will be a non-equilibrium stationary state.}

In obtaining the results (\ref{fluctuation theorem1}) and
(\ref{fluctuation theorem2}), we have assumed that the equations of
motion (\ref{closed macroscopic equation of motion}) allow the
time-reversed solution $x^{\dag}(t')$ for an arbitrary given
solution $x(t')$. However, as can be inferred from the expression
(\ref{macroscopic velocity}) for $\tilde{v}_i(x,t)$, although the
microscopic dynamics given by the Hamiltonian (\ref{total Hamiltonian})
is always time-reversible, the macroscopic
averaged dynamics (\ref{closed macroscopic equation of motion}) is
generally not time-reversible. In what follows,  we formulate the
condition for the ensemble to guarantee the time-reversibility of
the dynamics (\ref{closed macroscopic equation of motion}) for a
given time interval $[t_0,t]$; then the condition will give the
final form of the fluctuation theorem.

In order that the dynamics (\ref{closed macroscopic equation of
motion}) is time-reversible during the time interval $[t_0,t]$, the
equations of motion  should be invariant under the transformation
\begin{equation}
(x,t')  \rightarrow (x^{*}, t+t_{0}-t').
\end{equation}
Replacing $(x,t')$ with $(x^{*}, t+t_{0}-t')$ in (\ref{closed
macroscopic equation of motion}) and using (\ref{macroscopic
velocity}), we obtain
\begin{equation}\label{time-reversed equation}
\frac{dx^{*}_i(t')}{dt'} = -\int dx' \tilde{v}_i^{o}(x^{*},x')
\frac{\tilde{\rho} (x',t+t_{0}-t')}{\tilde{\rho}
(x^{*},t+t_{0}-t')},
\end{equation}
where, from (\ref{macroscopic velocity0}),
\begin{equation}\label{time reversed vo}
\tilde{v}_i^{o}(x^{*},x') =\frac{1}{\mathit{\Delta} t
{\mathit{\Delta} \mathit{\Gamma}}} \int_{0,x^{*}}^{\mathit{\Delta}}
ds  dy  \dot{x}_i(y)  e^{-i L(y) s}  \delta (y-x').
\end{equation}
In the above expression, using $L(y^{*})=-L(y)$ and (\ref{L relation for
delta}), and noting that $\tilde{v}_i^{o}(x,x')$ a slowly varying
function for $x$ and $x'$ as $\tilde{\rho}$ in the sense of
(\ref{condition1 for rho bar}) \footnote{In obtaining the second line
in (\ref{time-reversed v}), we
have used $\int_{p-\mathit{\Delta} p}^{p} \approx
\int_{p}^{p+ \mathit{\Delta} p}$ for $\mathit{\Delta} x \equiv
( \mathit{\Delta} q, \mathit{\Delta} p)$ as $\tilde{v}_i^{o}(x,x')$ is a
slowly varying function for $x$.}, we can write
{\setlength\arraycolsep{1.4pt}
\begin{eqnarray}\label{time-reversed v}
\tilde{v}_i^{o}(x^{*},x') & =& \frac{1}{\mathit{\Delta} t
{\mathit{\Delta} \mathit{\Gamma}}} \int_{x^{*},0}^{\mathit{\Delta}}
dy  ds   \dot{x}_i(y)  e^{-i L(y) s}  \delta (y-x') \nonumber \\
& = & \frac{1}{\mathit{\Delta} t  {\mathit{\Delta}
\mathit{\Gamma}}^2} \int_{x',x,0}^{\mathit{\Delta}} dy'  dy ds
\dot{x}_i(y^{*})  e^{-i L(y^{*}) s}
\delta (y^{*}-y') \nonumber \\
&=&\frac{1}{\mathit{\Delta} t  {\mathit{\Delta} \mathit{\Gamma}}^2}
\int_{x',x,0}^{\mathit{\Delta}} dy'  dy ds  \dot{x}_i(y^{*})
e^{-i L(y) (s- \mathit{\Delta} t)}  \delta (y-{y'}^{*}) \nonumber \\
&=&\frac{1}{\mathit{\Delta} t {\mathit{\Delta} \mathit{\Gamma}}^2}
\int_{x',x,0}^{\mathit{\Delta}} dy' dy ds \dot{x}_i(y^{*})
e^{-i L(y) s-i L({y'}^{*}) \mathit{\Delta} t}  \delta (y-{y'}^{*}) \nonumber \\
&=& \frac{1}{{\mathit{\Delta} \mathit{\Gamma}}}
\int_{x'}^{\mathit{\Delta}} dy'  e^{-i L(y'^{*}) \mathit{\Delta} t}
\tilde{v}_i^{o *}(x,{y'}^{*})
\end{eqnarray}
}with the definition
\begin{equation}\label{time reversed vo}
\tilde{v}_i^{o *}(x,x') \equiv \frac{1}{\mathit{\Delta} t
{\mathit{\Delta} \mathit{\Gamma}}} \int_{0,x}^{\mathit{\Delta}} ds
dy  \dot{x}_i(y^{*})  e^{-i L(y) s}  \delta (y-x').
\end{equation}

Let us consider the velocity field
obtained by replacing $\tilde{v}_i^{o}(x,x')$ in expression
(\ref{macroscopic velocity1}) with $\tilde{v}_i^{o *}(x,x')$, i.e.,
\begin{equation}\label{definition-of-v*}
\tilde{v}^{*}_i(x,t) \equiv \int dx'  \tilde{v}_i^{o *}(x,x'^{*})
\frac{\rho (x'^{*},t)}{\tilde{\rho} (x,t)}.
\end{equation}
Since $\tilde{v}^{*}_i(x,t)$ is different from $\tilde{v}_i(x,t)$
just in the sign of its components corresponding to $i \leq N$, the
velocity field  $\tilde{v}^{*}_i(x,t)$ is also a slowly varying function
for $t$ as $\tilde{v}_i(x,t)$ and $\tilde{\rho}(x,t)$. Therefore, if we
define $\mu_i (x,t) \equiv \tilde{\rho} (x,t)\tilde{v}^{*}_i(x,t)$,
the (time-reversed) probability current $\mu_i (x,t)$ varies slowly as
%
%
\begin{equation}\label{condition0 for rho v}
\mu_i (x, t + \mathit{\Delta} t) \approx \mu_i (x, t ) +
\frac{\partial  \mu_i (x, t )}{\partial t}  \mathit{\Delta} t.
\end{equation}
Also, in the same way as the process of obtaining (\ref{condition1 for
p}) and  (\ref{condition2 for p}), from the expression of $\mu_i$
given by (\ref{definition-of-v*}) and the
 Hermitian property of the operator $L(x,a)$, it follows that
\begin{equation}\label{condition1 for rho v}
\mu_i (x,t-\mathit{\Delta} t) = \int dx'  \rho (x'^{*},t) e^{-i
L(x'^{*}) \mathit{\Delta} t}  \tilde{v}_i^{o *}(x,x'^{*})
\end{equation}
and
\begin{equation}\label{condition2 for rho v}
\frac{\partial \mu_i (x,t)}{\partial t} = \int dx'  \rho (x'^{*},t)
i L(x'^{*})  \tilde{v}_i^{o *}(x,x'^{*}).
\end{equation}
Using (\ref{condition1 for rho v}) and (\ref{condition2 for rho v}),
the condition (\ref{condition0 for rho v}) becomes
\begin{equation}
\int dx'  \rho (x'^{*},t)  e^{-i L(x'^{*}) \mathit{\Delta} t}
\tilde{v}_i^{o *}(x,x'^{*}) \approx \int dx' \rho (x'^{*},t) \left [
1 - i  L(x'^{*}) \mathit{\Delta} t \right ] \tilde{v}_i^{o
*}(x,x'^{*}).
\end{equation}
And, with the substitution $\rho (x'^{*},t)=\tilde{\delta} (x'-y)$,
we have the approximation
\begin{equation}\label{condition for v0}
\frac{1}{\mathit{\Delta} \mathit{\Gamma}} \int_{y}^{\mathit{\Delta}}
dx'  e^{-i L(x'^{*}) \mathit{\Delta} t}  \tilde{v}_i^{o *}(x,x'^{*})
\approx \tilde{v}_i^{o *}(x,y^{*}) - \mathit{\Delta} t \cdot
\frac{i}{\mathit{\Delta} \mathit{\Gamma}} \int_{y}^{\mathit{\Delta}}
dx'  L(x'^{*})  \tilde{v}_i^{o *}(x,x'^{*}),
\end{equation}
which is sufficient for the description of the slow variation of
$\tilde{\rho}(x,t)$ as conditioned by (\ref{condition1 for rho
bar}).  Then, applying (\ref{condition for v0}) to
(\ref{time-reversed v}), the expression of
$\tilde{v}_i^{o}(x^{*},x')$ in macroscopic realm can be rewritten as
$\tilde{v}_i^{o}(x^{*},x')=\tilde{v}_i^{o *}(x,x'^{*})$, i.e.,
\begin{equation}\label{time-reversed v2}
\tilde{v}_i^{o}(x^{*},x') = \frac{1}{\mathit{\Delta} t
{\mathit{\Delta} \mathit{\Gamma}}} \int_{0,x}^{\mathit{\Delta}} ds
dy  \dot{x}_i(y^{*})  e^{-i L(y) s}  \delta (y-{x'}^{*}).
\end{equation}

Thus, substituting (\ref{time-reversed v2}) into (\ref{time-reversed
equation}) and considering that
$\dot{x}(y^{*})=(-\dot{q}(y),\dot{p}(y))$ for
$\dot{x}(y)=(\dot{q}(y),\dot{p}(y))$ and the definition
(\ref{macroscopic velocity0}), the transformed equations of motion
(\ref{time-reversed equation}) are identical to
{\setlength\arraycolsep{1.4pt}
\begin{eqnarray}\label{time-reversed equation2}
\frac{dx_i(t')}{dt'} & = & \int dx'  \tilde{v}_i^{o}(x,x'^{*})
\frac{\tilde{\rho} (x',t+t_{0}-t')}{\tilde{\rho} (x^{*},t+t_{0}-t')}
\nonumber\\
&=&\int dx'  \tilde{v}_i^{o}(x,x')  \frac{\tilde{\rho}
({x'}^{*},t+t_{0}-t')}{\tilde{\rho} (x^{*},t+t_{0}-t')}.
\end{eqnarray}
}On the other hand, by
(\ref{macroscopic velocity}) and (\ref{closed macroscopic equation
of motion}), we have
\begin{equation}\label{closed macroscopic equation of motion2}
\frac{dx_i(t')}{dt'} = \int dx'  \tilde{v}_i^{o}(x,x')
\frac{\tilde{\rho} (x',t')}{\tilde{\rho} (x,t')}.
\end{equation}
From the condition that (\ref{time-reversed equation2}) should
coincide with (\ref{closed macroscopic equation of motion2}) for
arbitrary $\tilde{v}_i^{o}(x,x')$, we obtain $\frac{\tilde{\rho}
({x'}^{*},t+t_{0}-t')}{\tilde{\rho} (x',t')}=\frac{\tilde{\rho}
(x^{*},t+t_{0}-t')}{\tilde{\rho} (x,t')}$, which holds for arbitrary
points $x$ and $x'$; thus, the condition for the time-reversibility
of the macroscopic dynamics (\ref{closed macroscopic equation of
motion}) is given by
\begin{equation}\label{time-reversibiltiy condition1}
\frac{\tilde{\rho} (x^{*},t+t_{0}-t')}{\tilde{\rho}
(x,t')}=C(t';t,t_0),
\end{equation}
where $C(t';t,t_0)$ is independent of $x$ and a function of $t'$
only, containing the parameters $t$ and $t_0$.
 In particular, in case of a stationary state, the condition (\ref{time-reversibiltiy
condition1}) is written as
\begin{equation}\label{time-reversibiltiy condition2}
\frac{\tilde{\rho} (x^{*})}{\tilde{\rho}
(x)}=C=\mspace{-1mu}\text{const}.
\end{equation}
Also, if we substitute $t'=t$ in (\ref{time-reversibiltiy
condition1}), it follows that
\begin{equation}\label{time-reversibiltiy condition3}
\frac{\tilde{\rho} (x^{*},t_{0})}{\tilde{\rho}
(x,t)}=C(t,t_0)=\mspace{-1mu}\text{const},
\end{equation}
where the last equality $C(t,t_0)=\mspace{-1mu}\text{const}$ means
that $C(t,t_0)$ is a fixed constant value for a given time interval
$[t_0,t]$.
Finally, using the above conditions (\ref{time-reversibiltiy
condition2}) and (\ref{time-reversibiltiy condition3}),
the fluctuation theorem \cite{Evans} follows from
(\ref{fluctuation theorem1}) and (\ref{fluctuation theorem2})
%
%
\begin{equation}\label{fluctuation theorem final}
\frac{p(x^{\dag}(t'); [t_0,t])}{p(x(t'); [t_0,t])} = C e^{-\delta
s}.
\end{equation}

Therefore, we have the conclusion that, for an arbitrary closed
Hamiltonian dynamical system with chaotic and ergodic properties, if
a trajectory $x(t)$ is observed in macroscopic measurement for
an arbitrary (equilibrium or non-equilibrium) thermodynamic process and
there exists non-zero probability to observe its time-reversed
trajectory $x^{\dag}(t)$ in the same process, the fluctuation
theorem (\ref{fluctuation theorem final}) always hold.

\section{Concluding Remarks}
In this paper, we have constructed the generalized non-equilibrium
extension of the entropy theory given by Gibbs and Einstein
\cite{Gibbs,Einstein}.  Also, based on the generalized entropy
theory, we have derived the fluctuation theorem.  The entropy
increment is directly connected with the presupposition [as is
represented by (\ref{fastmotion condition}) and (\ref{fastmotion
condition2})] that there are fast chaotic and ergodic motions
which cannot be described in a deterministic way under the given
macroscopic measurement system. In macroscopic observation, such
fast complex motions behave as an ensemble potential in phase space
(giving the interaction between ensemble elements), and resultantly,
the phase space volume is not conserved. The entropy increment
arises only for the imperfect observer who experiences the loss of
the deterministic dynamical information by measurement imperfection.
In this respect, the emergence of the irreversible physical time is
an intrinsic phenomenon to occur only for the imperfect observer by
that the observer constantly loses physical information.

\end{document}